\documentclass[12pt]{article}
\usepackage{graphicx}

\begin{document}
\title {Cosmological dark energy from the cosmic QCD phase transition 
and colour entanglement } 
\date{}
\maketitle
\vspace{-1in}
\author{\begin{center}{Shibaji Banerjee$^1$, 
Abhijit Bhattacharyya$^2$,
Sanjay K. Ghosh$^3$,\\
Ernst-Michael Ilgenfritz$^4$, 
Sibaji Raha$^{3}$
\footnote{Corresponding Author}
\footnote{Electronic Mail :
sibaji@bosemain.boseinst.ac.in}, 
Bikash Sinha$^5$, \\ 
Eiichi Takasugi$^6$ and Hiroshi Toki$^4$}
\end{center}
$^1$Physics Department, St. Xavier's College, 
30, Park Street, Kolkata 700016, INDIA \\
$^2$Physics Department, Scottish Church College, 1 \& 3, Urquhart Square, Kolkata 700006, INDIA \\
$^3$Physics Department, Bose Institute, 93/1, A. P. C. Road, 
Kolkata 700009, INDIA \\
$^4$Research Center for Nuclear Physics, Osaka University, Ibaraki, 
Osaka 567-0047, JAPAN \\
$^5$ Saha Institute of Nuclear Physics, 1/AF, Bidhannagar, Kolkata - 700 064, INDIA\\
$^6$ Graduate School of Physics, Osaka University, Toyonaka, Osaka 560-0043, JAPAN}
\begin{abstract}
Recent astrophysical observations indicate that the universe is composed of a large amount of dark energy (DE) responsible for an accelerated expansion of the universe, along with a sizeable amount of cold dark matter (CDM), responsible for structure formation. At present, the explanations for the origin or the nature of both CDM and DE seem to require ideas beyond the standard model of elementary particle interactions. Here, for the first time, we show that CDM and DE can arise entirely from the 
standard principles of strong interaction physics and quantum entanglement. Quantitative agreement with the present data obtains without the need for any adjustable parameters.
\end{abstract}

The past few decades have been momentous in the history of science insofar as several accurate astrophysical measurements have been carried out and cosmology, often considered a fair playground of exotic or fanciful ideas, has had to confront the reality of experiments. Based on the knowledge gleaned so far, the present consensus \cite{1,2,3} is that the standard 
model of cosmology, comprising the Big Bang and a flat universe is correct. The Big Bang Nucleosynthesis \cite{4} (BBN), which forms one of the basic tenets of the standard model, shows that baryons can at most contribute \( \Omega_B \) ( \( \equiv \frac{\rho_B}{\rho_c} \), \( \rho_c \) being the present value of closure density \( \sim 10^{-47} \) Gev$^4$ ) \( \sim \)0.04, whereas structure formation studies require that the total (cold) matter density should be \( \Omega_{CDM} \sim 0.23 \). Matter contributing to CDM is characterized by a dust-like equation of state, pressure \( p \approx \) 0 and energy density \( \rho > 0 \) and is responsible for clustering on galactic or supergalactic scales. Dark energy (DE), on the other hand, is smooth, with no clustering features at any scale. It is required to have an equation of state \( p = w\rho \) where \( w< 0 \) (ideally \( w = -1 \)), so that for a positive amount of dark energy, the resulting negative pressure would facilitate an accelerated expansion of the universe, evidence for which has recently become available from the redshift studies of type IA supernovae \cite{3,5}. 
For a flat universe \( \Omega \sim 1 \), \( \Omega_{DE} \sim 0.73 \) \cite{6} implies that \( \rho_{DE} \) today is of the order of \( 10^{-48} \) GeV$^4$.

There is no consensus within the community about the origin or the nature of CDM or DE. It has been argued that given the BBN limit of \( \Omega_B \sim 0.04 \), CDM cannot be baryonic; as a result, various exotic possibilities, all beyond the standard model \( SU(3)_c \times SU(2) \times U(1) \) of particle interactions, have been suggested. 
The situation is even more complicated for DE. The most natural explanation for DE would be a vacuum energy density, which {\it a priori} would have the correct equation of state (\( w = -1 \)). This possibility however is beset with the insurmountable difficulty that 
for any known (or conjectured) type of particle interaction, the vacuum energy density scale turns out to be many orders of magnitude larger than the present critical density. A trivial, but aesthetically displeasing, way out could be to {\it a priori} postulate a small cosmological constant. On the other hand, there exists in the recent literature a large number, too many to cite in this letter, of speculative suggestions, all substantially beyond the standard model. Apart from being non-standard, all these pictures require large (perhaps unacceptable) amounts of fine-tuning, as these authors themselves admit. It thus appears that one needs a second set of magic, in addition to the one needed for CDM, to explain DE. In this letter, we show that there can indeed be {\it Magic Without Magic}; there is no need to go beyond the standard model \( SU(3)_c \times SU(2) \times U(1) \) to understand the nature of CDM and DE. They can both arise from the same process of the cosmic quark-hadron phase transition occurring during the microsecond epoch after the Big Bang, provided we admit the existence of quantum entanglement for a strongly interacting system. Even though entanglement in quantum mechanics has been established both theoretically and experimentally, there is till date no evidence of such effects in strong interaction. We, however, postulate that being a quantum mechanical property, it would also occur in strong interactions too; more on this later on. Towards the end of this work, we shall {\it a posteriori} argue that it could have been anticipated that the Quantum Chromodynamic vacuum was the natural solution for the dark energy problem.

The role of phase transitions \cite{7} in the early universe has been recognized to be of paramount importance. In the strong interaction (Quantum Chromodynamics) sector, there is expected to be a distinct phase transition separating the confined (hadronic) phase from the deconfined (quark-gluon plasma) phase. In the early universe, this phase transition is predicted to occur during the microsecond epoch after the Big Bang. The order of this phase transition is at present an open issue. While it may be of second order or even a continuous transition (cross-over) in the laboratory, the cosmic phase transition could most likely be of first order, as has been forcefully argued \cite{8,9,10}. In what follows, we shall tacitly assume it to be of first order. We shall return to this issue later on. Other crucial {\it ans\"atze} in our scenario include 
that the universe is overall colour neutral at all times, which we believe nobody would argue with, and that baryogenesis is complete substantially before the QCD transition epoch and that the baryon number is an integer. The significance of these requirements would hopefully become obvious from what follows. Together with the assumption of quantum entanglement, the above picture is capable of explaining both dark matter and dark energy, as we shall argue below. 
\begin{figure}
\hspace{-3.63in}
\includegraphics[width=7.0in]{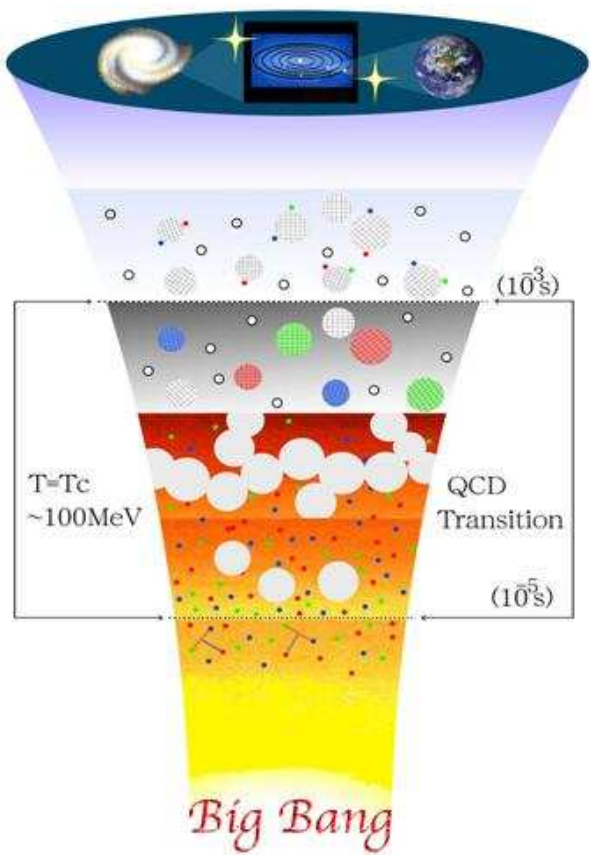} 
\vspace{-3.7in}
\caption{\label{fig:epsart} Schematic diagram of the first order cosmic quark-hadron 
transition. For explanation of the symbols, see text.}
\end{figure}

A first order phase transition can be described through a bubble nucleation scenario, as depicted in figure 1. At temperatures higher than the critical temperature \( T_c \), the coloured quarks and gluons are in a thermally equilibrated state in the perturbative vacuum (the quark-gluon plasma). The total colour of the universe is neutral (i.e., the total colour wave function of the universe is a singlet). It is strongly believed from QCD analysis that this thermally equilibrated phase of quarks and gluons exhibits collective plasma-like behaviour, referred to in the literature as the Quark-Gluon Plasma or QGP. Then, as \( T_c \) is reached and the phase transition starts, bubbles of the hadronic phase, denoted by grey circles, begin to appear in the quark-gluon plasma, grow in size and form an infinite chain of connected bubbles (the percolation process). At this stage, the ambient universe turns over to the hadronic phase in which the constituents (protons and neutrons) are denoted by small open circles. (Mesons are omitted from the diagram, as they do not carry any conserved quantum numbers, as opposed to the baryons.) Within this hadronic phase, the remaining high temperature quark phase gets trapped in large bubbles (denoted by big coloured and shaded circles). As is well known, this process is associated with a fluctuation in the temperature around \(T_c\); the bubbles of the hadronic phase can nucleate only when the temperature falls slightly below \(T_c\). The released latent heat raises the temperature again and so on. It is thus fair to assume that the temperature of the universe remains around \(T_c\) at least upto percolation.

Witten \cite{9} argued some time ago that the net baryon number contained in these Trapped False Vacuum Domains (TFVD) could be many orders of magnitude larger than that in the normal hadronic phase and they could constitute the absolute ground state of strongly interacting matter. It has been shown \cite{11} that if these TFVDs possess baryon number in excess of \( 10^{42-44} \), they would be stable on cosmological time scales and would form the so-called strange quark nuggets (SQN). In such situations, they would have spatial radii $\sim$ 1 m, while their spatial separation would be $\sim$ 300 m. TFVDs with baryon number less than this would evaporate into normal baryons quite rapidly, much before BBN starts. (To distinguish the baryon number contained in SQNs from that participating in BBN, we denote the SQN matter as quasibaryonic.) The SQNs could evolve \cite{12} to primordial structures of approximate solar mass, which should manifest themselves as the Massive Compact Halo Objects (MACHO), discovered \cite{13,14} through gravitational microlensing in the Milky Way Halo in the direction of the Large Magellanic Cloud (LMC). If all the CDM is taken to be in the form of MACHOs, then their estimated abundance in the Milky Way Halo is seen to correspond to the observed optical depth \cite{12}.

In all these considerations, the explicit role of colour, the fundamental charge of strong interaction physics, has been glossed over. It has been tacitly assumed that in a many-body system of quarks and gluons, colour is averaged over, leaving only a statistical degeneracy factor for thermodynamic quantities. In this letter, we argue that such simplification may have led us to overlook a fundamentally important aspect of strong interaction physics in cosmology. 

Considering the heterogeneous readership at whom this letter is aimed, it may be worthwhile to elaborate on the situation in some detail. In the QGP, all colour charges are neutralised within the corresponding Debye length, which turns out to be \( \sim {1 \over{g_s(T) T}} \), where \(g_s\) is the strong coupling constant. If this Debye length is smaller than a typical hadronic radius, then hadrons cannot exist as bound states of coloured objects. The lifetime of the QGP may be roughly estimated by the temperature when the Debye screening length becomes larger than the typical hadronic radius, and formation of hadrons as bound states of coloured objects becomes possible. 

Since the size of the universe is many orders of magnitude larger than the Debye length, the requirement of overall colour neutrality is trivially satisfied.  Another condition required for the existence of the QGP would be the occurrence of sufficient number of colour charges within the volume characterised by the Debye length; otherwise the collective behaviour responsible for the screening would not be possible. For the cosmic QGP, these conditions are satisfied till temperatures \( \sim \) 100 - 200 MeV, the order of magnitude for the critical temperature for QCD. 

In the light of the above, one would argue that all the physics of the QGP is contained within a Debye length. A quick estimate would show that upto
\(T_c\), the Debye length is less than a fermi and the total number of colour charges (including quarks, antiquarks and gluons) within the corresponding volume is greater than 10. Note, however, our emphasis on the second ansatz above, that the baryogenesis has already taken place much before \(T_c\) is reached and the value of \({{n_b} \over {n_\gamma}} \sim 10^{-10} \) has already been established. Firstly, it has to be realised that this baryon number is carried in quarks at this epoch, so that the net quark number (\( N_q - N_{\bar{q}} \)) in the universe 
at all times is an exact multiple of 3. The net quark number within a Debye volume then turns out to be \(\sim 10^{-9}\). Thus, to ensure overall colour neutrality and an integer baryon number, one must admit of long-range correlations beyond the Debye length in the QGP, the quantum entanglement property \cite{15}, which was identified as an essential feature of quantum mechanics by Schr\"odinger in as early as 1927 but was experimentally established only recently. 

Let us then consider the process of the cosmic quark-hadron phase transition from the quantum mechanical standpoint of colour confinement. As already mentioned, the colour wave function of the entire universe prior to the phase transition must be a singlet, which means it cannot be factorized into constituent product states; the wave functions of all coloured objects are completely entangled \cite{15} in a quantum mechanical sense. (It may be mentioned at this juncture that the requirement of the entangled colour singlet wave function of the universe also ensures the integer baryon number for any non-zero baryon chemical potential. A quick group theoretic analysis would show that one must have net quark numbers in exact multiples of 3, irrespective of the number of quarks, antiquarks and gluons.) In such a situation, the universe is characterized by a vacuum energy corresponding to the perturbative vacuum of Quantum Chromodynamics (QCD). As the phase transition proceeds, locally colour neutral configurations (hadrons) arise, resulting in gradual decoherence of the entangled colour wave function of the entire universe. Note that the coloured objects within the hadrons are entangled among themselves but not with those in the rest of the universe. This amounts to a proportionate reduction in the perturbative vacuum energy density, which goes into providing the latent heat of the transition, or in other words, the mass and the kinetic energy of the particles in the non-perturbative (hadronic) phase. (It should be mentioned here that the vacuum energy of the non-perturbative phase of QCD is taken to be zero; more on this later.) In the quantum mechanical sense of entangled wave functions, the end of the quark-hadron transition would correspond to complete decoherence of the colour wave function of the universe; the entire vacuum energy would disappear as the perturbative vacuum would be replaced by the non-perturbative vacuum. It must be emphasized that this is an example of decoherence through a classical mechanism; the phase transition leads to a different quantum state with different entanglement property.

Combining these observations with the formation of TFVDs as discussed above, it is obvious that in order for the TFVDs to be stable physical objects, they must be colour neutral. This is synonymous with the requirement that they all have integer baryon numbers, i.e., at the moment of formation each TFVD has net quark numbers in exact multiples of 3. For a statistical process, this is, obviously, most unlikely and consequently, most of the TFVDs would have some residual colour at the percolation time, as shown in fig. 1. (This is not at all inconsistent with the screening of colour within a Debye length, even if the size of the TFVD is much larger than the Debye length. It is well known from the study of the elctromagnetic plasma that whenever a boundary occurs in the plasma, the charge neutrality condition is violated over a small length of the order of the Debye length at the boundary, the plasma sheath effect.) Then, on the way to becoming colour singlet (shaded large grey circles), they would each have to shed one or two coloured quarks, as depicted.

Thus, the end of the cosmic QCD phase transition corresponds to a situation where there would be a few coloured quarks, separated by spacelike distances. It has to be noted that such a large separation, apparently against the dictates of QCD, is by no means unphysical. The separation of coloured TFVDs occurs at the temperature \( T_c \), when the effective string tension is zero, so that there does not exist any long range force. By the time the TFVDs have evolved into colour neutral configurations, releasing the few orphan coloured quarks in their immediate vicinity, the spatial separation between these quarks is already too large to allow strings to develop between them; see below. (It should be emphasized here that such a situation could not occur in the laboratory searches for quark-gluon plasma through energetic heavy ion collisions. There the spatial extent of the system is of the order of a few fermi and the reaction takes place on strong interaction time scales.) Therefore, the orphan quarks must remain in isolation. In terms of the quantum entanglement and decoherence of the colour wave function, this would then mean that their colour wave functions must still remain entangled and {\it a corresponding amount of the perturbative vacuum energy would persist in the universe}. In this sense, the orphan quarks are definitely not in asymptotic states and no violation of colour confinement is involved. If we naively assume that the entanglement among these orphan quarks imply collective behaviour like a plasma, then one can estimate the corresponding Debye screening length for this very dilute system of quarks, which would still be governed primarily by the temperature. At temperatures of \( \sim \) 100 MeV, the scale would be a few fm, much smaller than the mutual separation between the orphan quarks; thus the formation of bound states of orphan quarks would be impossible. 

There does not exist any way to calculate this persisting, or even the full, perturbative vacuum energy, from first principles in QCD. For the latter quantity, one may adopt the phenomenological Bag model \cite{16} of confinement, where the Bag parameter B (\(\sim (145 MeV)^4 \)) is the measure of the difference between the perturbative and the non-perturbative vacua. Thus we can assume that at the beginning of the phase transition, the universe starts out with a vacuum energy density B, which gradually decreases with increasing decoherence of the entangled colour wave function. A natural thermodynamic measure of the amount of entanglement during the phase transition could be the volume fraction 
(\( f_q \equiv V_{colour}/V_{total} \)) of the coloured degrees of freedom; at the beginning, \(f_q\) is unity, indicating complete entanglement, while at the end, very small but finite entanglement corresponds to a tiny but non-zero \(f_q\) due to the coloured quarks. Accordingly, the amount of perturbative vacuum energy density in the universe at any time is the energy density B times the instantaneous value of \(f_q\); within the scenario discussed above, the remnant perturbative vacuum energy at the end of the QCD transition would just be B \( \times f_{q,O}\), where \(f_{q,O}\) is due solely to the orphan quarks. An order of magnitude estimate for \(f_{q,O}\) can be carried out in the following straightforward manner. On the average, each TFVD is associated with 1 orphan quark so that the number \( N_{q,O} \) of orphan quarks within the horizon volume at any time is about the same as the number \( N_{TFVD} \) of TFVDs therein. It is well known from the study of percolating systems \cite{stauffer} that percolation is characterized by a critical volume fraction \(f_c \sim \) 0.3 of the high temperature phase. In the present case, this would require \(f_q \) in the form of TFVD-s to be \( \sim \) 0.3. Following the ansatz of Witten \cite{9} that the most likely length scale for a TFVD is a few cm, one can estimate \(N_{TFVD}\) (and hence \(N_{q,O}\)) within the horizon at the percolation time of about 100 \(\mu\)sec \cite{8} to be about 10\(^{18-20}\). The inter-TFVD separation comes out to be \( \sim \) 0.01 cm at that time. (It is obvious that the orphan quarks, separated by distances of 0.01 cm, cannot develop colour strings between them, even if there is some non-zero string tension generated at temperatures slightly lower than \(T_c\).) Then, if we naively associate an effective radius of \(\sim 10^{-14} cm\) (estimated from \( \sigma_{qq} = \frac{1}{9} \sigma_{pp}~;~~ \sigma_{pp} \sim 20 mb \) ) with each orphan quark, we obtain \( f_{q,O} \sim N_{q,O} \times (v_{q,O}/V_{total}) \sim 10^{-42} - 10^{-44}\) (where $v_{q,O}$ is the effective volume of an orphan quark), so that the residual pQCD vacuum energy comes out to be in the range 10\(^{-46}\) to 10\(^{-48} GeV^4\), just the amount of DE. 

Admittedly, the preceding estimate is somewhat crude; nevertheless, the simple and natural assumptions that have led to this value ensure that it cannot be very wrong. An alternate treatment within a slightly more microscopic model also leads to the same estimate, as we show now. 

The problem of finding an effective inter-quark potential at large \( r \) from QCD is well known. In the spirit of the phenomenological string model for a quark-antiquark system (mesons), one expects a linear potential. It was shown by Richardson \cite{richardson} some time ago that one can postulate an effective coupling constant of QCD which gives the usual leading order form at high \( Q^2 \) 
(small \(r\)) but can be carried over to small \(Q^2\) (large \(r\)) limit without going through the singularity. The potential derived from such prescription yields a linear potential in the large \(r\) limit. This "standard" form of interaction is, unfortunately, not applicable in the context of the present analysis since here we are interested in the dynamics of interaction in a dilute many body quark system, where there may not be any antiquarks and certainly no mesons. It may however be recalled that a phenomenological prescription for incorporating the effect of confinement in such systems was proposed many years ago \cite{fowler}, through the use of a density dependent effective quark mass \(m_q \sim 1 /n_q \) in the low density sector. This prescription has been used extensively by a number of authors to study the properties of quark matter. In addition, Pl\"umer and Raha \cite{pluemer} showed quite some years ago that such a behaviour of the effective mass was consistent with a running coupling constant which behaved like \( \alpha_s \sim 1/log(1+Q^4/\Lambda^4)\). Note that this expression agrees with the familiar perturbative form (leading order) for large \(Q^2\), just like the Richardson case but has a very different large \(r\) behaviour. 
Using this \(\alpha_s\), one can find the effective inter-quark potential by taking the Fourier transform of \(V(q) = \alpha_s(q^2)/q^2\) {\it a la} Richardson \cite{richardson}. This gives
\begin{equation}
V(r) \sim \Lambda \left( (\Lambda r)^3 - \frac{12}{\Lambda r} \right) 
\end{equation}
For large \(r\), only the cubic term needs to be retained. This (cubic) potential, rather than the linear one, is the most natural form of the potential to use in the case of a dilute many body system of quarks.

The inter-quark separation \(r\) can be estimated as follows:
\begin{eqnarray}
n_{q,O} \equiv N_{q,O}/V_H = (3/4 \pi) N_{q,O}/R_H^3 \\ 
r = \left(\frac{3}{4 \pi} n_{q,O} \right)^{1/3} = R_H/N_{q,O}^{1/3} 
\end{eqnarray}

Then, the potential energy density for this inter-quark interaction is
\begin{equation}
\rho_V = \frac{1}{2} n_{q,O} V(r) \sim \frac{3}{8 \pi} \Lambda^4
\end{equation}
Note that this is independent of density and hence, time; once it comes into being, it will retain the same value through the subsequent evolution of the universe and can indeed be the ideal candidate for dark energy. All we need to do is to fix the value of \(\Lambda\), which would be appropriate for this dilute system of quarks. A natural measure of \(\Lambda\) would be the inverse of the length scale characterising this dilute system of quarks. For hadronic bags or hard QCD processes, the length scale is of the order of Fermi, which gives \(\Lambda \sim \) 100 MeV. In our case, the appropriate length scale should be the size of the smallest TFVDs, as Kodama, Sasaki and Sato \cite{kodama} argued many years ago. For the stable SQNs, with baryon density \(\sim 10^{38} cm^{-3}\), we had calculated the length scale for smallest TFVD \cite{abhi-prd} to be several cm. Now the baryon number density in the universe during the microsecond epoch is \( \sim 10^{30}  cm^{-3} \). So, for smaller TFVDs, whose baryon density at the time of formation would be \(\sim 10^{30} cm^{-3}\) (corresponding to \(T_c \sim\) 100 MeV), the appropriate length scale must be smaller by about two orders of magnitude, \(\sim\) 0.01 cm. (It is no surprise that the inter-TFVD separation, the other possible length scale in the problem, as estimated above comes out to be of the same order of magnitude.) Then, the corresponding \(\Lambda\) would be \(\sim 10^{-12}\) GeV, so that \(\rho_V \sim 10^{-48} GeV^4\), just the value required for dark energy.

Even though the DE component appears during the microsecond era in the history of the universe, it remains negligible in comparison to the matter density for most of the history. Since matter density decreases as \( R^{-3} \) (R being the scale size), while DE density remains constant, the latter can become dominant only at very late times (\( z \sim 0.17 \)) and thus would not affect the galaxy formation scenarios to any extent.

The density of the orphan quarks in the present universe is exceedingly small; compared to \( \sim 10^{77} \) baryons which took part in BBN, there would be \( 10^{44-45} \) orphan quarks. Their flux in the cosmic rays would be negligibly small; non-observance of fractionally charged objects is thus not a detractor. A definitive detection of fractional electric charge would certainly clinch the issue, but as mentioned above, the flux of orphan quarks seems to make it highly urealistic. A conclusive detection of strange quark matter in the cosmic ray flux, however, would go a long way in providing support to the idea. Many such experiments are currently under way.

The picture presented here relies on a putative first order transition, although any transition resulting in finite size fluctuations may suffice. This issue requires further investigation.

{\it A posteriori}, it is tempting to mention that it should perhaps have been anticipated that if the cosmological constant does arise from a vacuum energy, then the QCD vacuum is the most natural candidate. It is a known (and accepted) lore of (renormalisable) quantum field theories that the divergent vacuum energy density is renormalised to the physical parameters of the theory. Within all the field theories now in vogue, it is only the QCD which has a distinct perturbative and a non-perturbative vacuum, which are separated by a finite energy density, irrespective of the renormalisation prescription. (That we are unable to estimate this as yet from first principles in QCD is a technical shortcoming, not a conceptual one.) It is thus most plausible that even if suitable renormalisation prescriptions remove all vacuum energy densities, the finite part of the perturbative vacuum energy density of QCD should survive and play the role of the cosmological constant.
  
We therefore conclude by reiterating that the emergence of both CDM and DE from the same mechanism {\it entirely within the standard model of particle interactions}, and that too in quantitatively correct amounts without any fine-tuning of parameters, is nothing short of remarkable. The picture presented here satisfies, in our opinion, the most stringent criterion of a scientific theory, that of naturalness. The strangest property of the universe, as it is said, is how simple it is.

The authors are indebted to Prof. Anthony J. Leggett for his critical comments on an earlier version of the manuscript. SR would also like to thank the Research Center for Nuclear Physics, Osaka University for their warm hospitality during his sojourn there.

\end{document}